\documentclass[amssymb,amsmath,aps,showpacs,floatfix,nofootinbib,showpacs,12pt]{revtex4}
\usepackage{epsfig}
\usepackage{color}
\usepackage{graphicx}

\begin{document}
\title{The Friedberg-Lee Symmetry and Minimal Seesaw Model
}

\author{$^{1,2}$Xiao-Gang He and $^{1,3}$Wei Liao}
\affiliation{ $^1$Center for High Energy Physics, Peking University, Beijing\\
$^2$Department of Physics and Center for Theoretical
Sciences, National Taiwan University, Taipei\\
$^3$ Institute of Modern Physics, East China University of Science
and Technology, Shanghai}

\begin{abstract}
The Friedberg-Lee (FL) symmetry is generated by a transformation of
a fermionic field $q$ to $q + \xi z$. This symmetry puts very
restrictive constraints on allowed terms in a Lagrangian. Applying
this symmetry to $N$ fermionic fields, we find that the number of
independent fields is reduced to $N-1$ if the fields have gauge
interaction or the transformation is a local one. Using this
property, we find that a seesaw model originally with three
generations of left- and right-handed neutrinos, with the
left-handed neutrinos unaffected but the right-handed neutrinos
transformed under the local FL translation, is reduced to an
effective theory of minimal seesaw which has only two right-handed
neutrinos. The symmetry predicts that one of the light neutrino
masses must be zero.
\end{abstract}

\pacs{14.60.Pq, 13.15.+g}

\maketitle

In trying to understand the properties of neutrinos, Friedberg and
Lee~\cite{Friedberg:2006it} proposed a symmetry translating a
fermionic field $q$ to $q + \xi z$ where $z$ is an element of
Grassmann algebra and $\xi$ is a complex number. We will call this
symmetry the Friedberg-Lee (FL) symmetry. Various applications of
the FL symmetries have been
studied~\cite{Friedberg:2007ba,Jarlskog:2007qy,Lee:2008zzh,xing1,xing2,
Huang:2008ri,Jarlskog:2008zf,Friedberg:2008tf,Luo:2008yc,
Friedberg:2008ja,Araki:2008xq,Araki:2009ds,BarShalom:2009sk,
Friedberg:2009fb,Araki:2009kp,Chan:2009du}. In this Letter we
further study some properties of the FL symmetry being a global or a
local symmetry, and apply to neutrino seesaw models. We find that
applying the FL symmetry to the whole Lagrangian is dramatically
different than applying the same symmetry only to terms related to
fermion masses. In the latter case the FL symmetry along a certain
direction implies a zero mass eigenstate of fermions, but in the
former it implies complete decoupling of the same field in the
theory if the fermionic fields have gauge interaction or the FL
transformation is local. That is, applying the FL symmetry to $N$
fermionic fields, we find that the number of independent fields is
reduced to $N-1$. Using this property, we find that a seesaw model
originally with three generations of left- and right-handed
neutrinos, with the left-handed neutrinos unaffected but the
right-handed neutrinos transformed under the local FL translation,
is reduced to an effective theory of minimal seesaw which has only
two right-handed neutrinos.

\vspace{0.3cm}
\noindent
{\bf The FL symmetry and number of independent fields}

 Assuming that there are $N$ generations of fermion fields $N_R^i$
 with certain gauge charges. Under a FL transformation
 $N_R^i$ transform as
 \begin{eqnarray}
  N^i_R \to N^i_R + \xi_i z, \label{FL1}
 \end{eqnarray}
 with $z$ an element of the Grassmann algebra, anti-commuting with the
 field operator $N^i_R$. As an element of the Grassmann algebra, $z$ can
 be space-time independent or space-time dependent. $z=(z_1,z_2)^T$,
 with $z_\alpha (\alpha=1,2)$ two Grassmann numbers, is a
 two-component spinor if using two-component theory describing
 fermionic field.  $z$ is a four-component spinor if using four-component theory
 describing fermionic field. $\xi_i(i=1,\cdots, N)$ is a particular
 set of c-numbers, similar to that used in Ref.~\cite{Friedberg:2007ba} for quarks.
 In-equivalent choices of $\xi_i$ say that fermionic fields are translated
 in different directions in $N$-dimensional space of $(N^1_R,\cdots,
 N^N_R)$. With a particular set of $\xi_i$  we
 implement the FL translation of $N^i_R$ only along a specific
 direction described by a set of $\xi_i$ following Ref.~\cite{Friedberg:2007ba}.

For a theory having only these $N$ fermionic fields, one can write the
renormalizable Lagrangian as the following
\begin{eqnarray}
 {\cal L}_R = \gamma_{ij} \bar N_R^i \gamma_\mu (i D^\mu N^j_R)
 - {1\over 2} \left [ m_{ij} \bar N_R^{ic} N^j_R + H.C. \right ]
\;,\label{original}
\end{eqnarray}
where $i,j=1,\cdot\cdot\cdot,N$ and summation over repeated indices is
assumed. $\gamma_{ij}$ and $m_{ij}$ are Hermitian and symmetric
$N\times N$ matrices, respectively. $N^c_R$ is the charge conjugated
field of $N_R$. $D^\mu = \partial^\mu + i g A^\mu$ is the covariant
derivative, $A^\mu$ is the gauge field and $g$ is the gauge coupling.

Under the transformation Eq. (\ref{FL1}) the Lagrangian transforms
as
 \begin{eqnarray}
 {\cal L}_R && \to {\cal L}_R+ \gamma_{ij} \xi_j\bar N_R^i
 \gamma_\mu (i D^\mu  z)  +\gamma_{ij} \xi^*_i\bar z
 \gamma_\mu  (i D^\mu N_R^j) + \gamma_{ij}\xi^*_i \xi_j \bar z \gamma_\mu ( iD^\mu  z) \nonumber\\
 && -{1\over 2} \left [ m_{ij}\xi_j  N_R^{ic} z + m_{ij} \xi_i^* z^c N^j_R
 + m_{ij} \xi^*_i \xi_j z^c z + H.C. \right  ]
 \end{eqnarray}

Requiring that the Lagrangian ${\cal L}_R$ to be invariant under the
FL symmetry, for the case with $g\neq 0$, implies
\begin{eqnarray}
\gamma_{ij}\xi_j = 0\;,\;\;\;\;m_{ij} \xi_j = 0\;.\label{symm}
\end{eqnarray}
Both equations imply that the linear combination $N_R^0 = \sum_{i}^N
\xi_i N^i_R/\sqrt{\sum_j \xi^*_j \xi_j}$ is an eigenvector
corresponding to zero eigenvalues for $\gamma_{ij}$ and $m_{ij}$
matrices. It has been pointed
out~\cite{Jarlskog:2007qy,Jarlskog:2008zf,Chan:2009du} that if one
requires the above equations to be true for an arbitrary set of
parameters $\xi_i$ (a generic FL symmetry), then there are N number
of zero eigenvalues, that is, $m_{ij}$ must be zero. As have been
mentioned before that we follow Ref.~\cite{Friedberg:2007ba} to
choose FL invariance along a particular direction in $\xi_i$
parameter space. Therefore there is only one zero eigenvalue for
$m_{ij}$ and also for $\gamma_{ij}$. Note that the zero eigenvalues
in both $\gamma_{ij}$ and $m_{ij}$ have the same eigenvector does
not mean that the $\gamma_{ij}$ and $m_{ij}$ can be, in general,
simultaneously diagonalized by unitary transformations in the form
$V^\dagger \gamma V = \hat \gamma$ and $V^T m V = \hat m$. Here
$\hat \gamma$ and $\hat m$ are diagonal matrices.

If $g=0$, applicable if $N_R$ is right-handed neutrino, depending on
whether the FL transformation is global or local, there are
different implications. If the FL is a global symmetry, that is $z$
is independent of space-time which leads to $\partial^\mu z = 0$,
the kinetic energy terms are invariant up to terms proportional to
total derivatives. There is no constraint on the form of
$\gamma_{ij}$. However, if the transformation is local as discussed
in Ref.~\cite{Huang:2008ri}, that is $\partial^\mu z \neq 0$, the
kinetic terms are not invariant under the FL transformation unless
$\gamma_{ij} \xi_j = 0$.

If one only applies the FL symmetry to the mass term, regardless
whether the FL is global or local, one predicts a zero
eigenmass~\cite{Friedberg:2006it,Friedberg:2007ba,Jarlskog:2007qy,Lee:2008zzh,xing1,xing2,
Huang:2008ri,Jarlskog:2008zf,Friedberg:2008tf,Luo:2008yc,
Friedberg:2008ja,Araki:2008xq,Araki:2009ds,BarShalom:2009sk,
Friedberg:2009fb,Araki:2009kp,Chan:2009du}. If one applies the FL
symmetry to the whole Lagrangian $L$, the consequences are
different. Taking the latter as requirement for the Lagrangian, we
find that, if the fermionic fields have gauge interaction or the FL
transformation is local, the eigenvector corresponding to the zero
eigenvalues of $\gamma$ and $m$ matrices completely decouples from
the theory. To see this let us work in the basis where $\gamma$ is
in a diagonalized form,
\begin{eqnarray}
\hat \gamma = \left (
\begin{array}{lllll}
\gamma_1&0&\cdot \cdot\cdot&0&0\\0&\gamma_2&\cdot\cdot \cdot&0
&0\\\cdot\cdot\cdot&\cdot\cdot\cdot&\gamma_{i}
&\cdot\cdot\cdot&\cdot\cdot\cdot\\0&0&\cdot\cdot\cdot&\gamma_{(N-1)}&0\\0&0&\cdot\cdot\cdot&0&0
\end{array} \right )\;,
\end{eqnarray}
The $m$ matrix in this basis must be able to be written, due to Eq.
(\ref{symm}), in the following form
\begin{eqnarray}
m = \left ( \begin{array}{lllll} m_{11}&m_{12}&\cdot\cdot\cdot&m_{1
(N-1)}&0\\m_{12}&m_{22}&\cdot\cdot\cdot&m_{2
(N-1)}&0\\\cdot\cdot\cdot&\cdot\cdot\cdot&\cdot\cdot\cdot&\cdot\cdot\cdot&\cdot\cdot\cdot\\m_{1
(N-1) }&m_{2 (N-1)}&\cdot\cdot\cdot&m_{(N-1)
(N-1)}&0\\0&0&\cdot\cdot\cdot&0&0
\end{array} \right )\;. \label{m-matrix}
\end{eqnarray}

This implies that when writing in eigenvectors of $\gamma$, the
linear combination $N_R^0$ does not show up anywhere in the
Lagrangian. Assuming that the eigenvectors correspond to the
non-zero eigenvalues $\gamma_{i}$ are $\nu^{\prime 1},\nu^{\prime
2}_R,\cdots, \nu^{\prime N-1}$, Lagrangian in Eq. (\ref{original})
is reduced to
\begin{eqnarray}
{\cal L}_R = \gamma_i \bar \nu_R^{\prime i} \gamma_\mu (i D^\mu
\nu_R^{\prime i})  - {1\over 2} \left [\tilde m'_{ij} \bar
\nu_R^{\prime ic} \nu^{\prime j}_R + H.C. \right ]\;,
\end{eqnarray}
where $\tilde m'$ matrix is the left-upper corner $(N-1)\times
(N-1)$ matrix in Eq. (\ref{m-matrix}).

By a re-scaling of the field $\nu^i_R = \sqrt{\gamma_i} \nu^{\prime
i}_R$, one can  write the Lagrangian in the usual form
\begin{eqnarray}
 {\cal L}_R = \bar \nu_R^{i} \gamma_\mu
(i D^\mu \nu_R^{i}) -  {1\over 2} \left [\tilde m_{ij} \bar
\nu_R^{ic} \nu^{j}_R + H.C. \right ]\;.
\end{eqnarray}
One can further diagonalize $\tilde m = U^T \hat m U$, with $U$
a unitary matrix, to obtain normalized mass eigenstates $\nu_R^{m i}
= U_{ij} \nu^j_R$. We finally have
\begin{eqnarray}
 {\cal L}_R = \bar \nu_R^{mi} \gamma_\mu
(i D^\mu \nu_R^{mi}) -  {1\over 2} \left [ \hat m_{i} \bar
\nu_R^{mic} \nu^{mi}_R + H.C. \right ]\;.\label{ff}
\end{eqnarray}

The above is a Lagrangian for $N-1$ independent fields. Starting
with $N$ fermionic fields, after imposing the FL symmetry introduced
in Eq. (\ref{FL1}), the number of independent fields has been
reduced by one if the fermionic fields have gauge interaction or the
FL transformation is local. If the fermionic fields have no gauge
interaction and the FL transformation is global, the number of
independent fields is not affected.

One can understand the reduction of the number of fields in a
different way as the following. One can build a Lagrangian which is
invariant under the FL transformation by using all independent
combinations of $N^i_R$ which do not transform under the FL symmetry
as building block. We note that $\xi_j N^i_R - \xi_i N^j_R$ is
manifestly invariant under the FL transformation. They should be
naturally used to build $L$. Because this construction is taking a
difference of two fields, out of $N$ fields  only $N-1$ such
differences are independent. For example, if one takes $q_j = \xi_j
N^1_R - \xi_1 N^j_R$ as the $N-1$ independent ones, $\xi_j N^2_R -
\xi_2 N^j_R$ can be expressed as $(\xi_2 q_j - \xi_j q_2)/\xi_1$.
Similarly for other combinations. Imposing the FL symmetry, Eq.
(\ref{FL1}), to a theory with $N$ number of fields, only $N-1$ are
dynamic fields which are the real physical degrees of freedom in the
theory, not all the $N$ number of the field. The one drops out of
the theory is $N^0_R$ which is the linear combination of $N^i_R$ in
accordance with the FL translation introduced in Eq. (\ref{FL1}) .
In another words, to have a theory having $N$ number of dynamic
fermion fields with a FL symmetry given in Eq. (\ref{FL1}), one must
start with a theory containing $N+1$ fields.
 \vspace{0.3cm}

\noindent {\bf The FL symmetry and Seesaw Models}

We now study seesaw models with FL symmetry. The simplest seesaw
model~\cite{seesaw1} is the minimal standard model (SM) with
additional right-handed neutrinos $N_R^i$. Experimentally there are
three light neutrinos with SM charged current interaction, a
successful model for neutrinos must have three left-handed minimal
SM lepton doublets. The number of right-handed neutrinos can, in
principle, be different than their left-handed ones. But there
should be at least two right-handed neutrinos in order to satisfy
experimental constraint that two of the light neutrinos are massive.
This is the so-called minimal seesaw model. It is sometimes also
called the 3+2 seesaw model. This model has some interesting
consequences~\cite{Frampton:2002qc,Barger:2003gt,xing10}, such as a
zero mass light neutrino and possible connection of CP violating
source for baryon asymmetry and low energy CP violating phases. A
more symmetric model is the 3+3 seesaw model in which both the left-
and right-handed are three generations. We find that a local FL
symmetry can make a passage from a 3+3 seesaw model to a 3+2 minimal
seesaw model making the theory with more predictive power. In the
following we show this in details.

Particles relevant to our discussions are the three generations of
left-handed lepton doublets $L^i_{L} = (\nu^i_L, e^i_L)^T$, the
three right-handed neutrino singlets $N^i_R$, and the Higgs doublet
$H = (H^0/\sqrt{2}, H^-)^T$. The transformation properties of these
fields are as follows. The left-handed leptons $L_L$  and Higgs
boson $H$ do not transform under a local FL transformation, but the
right-handed neutrinos $N^i_R$ do:
 \begin{eqnarray}
 L_L \to L_L,~~ H \to H,~~ N_R^i \to N^i_R + \xi_i z. \label{FL2}
 \end{eqnarray}

As have been seen from our previous discussions that the local FL symmetry
restricts the forms of allowed terms in the Lagrangian, we should pay
special attentions to the fields transforming non-trivially under
the FL symmetry. To this end we write all renormalizable terms
involving $N^i_R$ in the following for detailed analysis,
\begin{eqnarray}
{\cal L} = \gamma_{ij}\bar N_R^i \gamma_\mu (i
\partial^\mu N_R^j) - {1\over 2} \left [ m_{ij} \bar N_R^{ic} N^{j}_R  + 2 Y'_{ij}
\bar L^i_L H N^j_R + H.C. \right ]\;.
\end{eqnarray}
Again, $\gamma$ is Hermitian, $m$ is symmetric. But there is no
constraint on the form of $Y'$ before applying the FL symmetry.

The requirement that ${\cal L}$ being invariant under a local FL
symmetry constrains the form of $\gamma_{ij}$, $m_{ij}$ and
$Y'_{ij}$. Similarly to Eq. (\ref{symm}) we have
\begin{eqnarray}
\gamma_{ij} \xi_j = 0\;,\;\;\;\;m_{ij}\xi_j =
0\;,\;\;\;\;Y'_{ij}\xi_j = 0\;. \label{constraint2}
\end{eqnarray}
In general the matrix $Y'$ can be written in the following form
\begin{eqnarray}
 Y'= y'_1 u_1 v_1^\dagger +y'_2 u_2 v_2^\dagger + y'_3 u_3 v_3^\dagger,
 \label{Yukaw1}
\end{eqnarray}
 where $u_i$ are eigenvectors of $Y' Y'^\dagger$and $v_i$ are
 eigenvectors of $Y'^\dagger Y'$. The constraint on $Y'$ in
 Eq. (\ref{constraint2}) implies that $v_3=(\xi_1,\xi_2,\xi_3)^T$ and $y'_3=0$.
 $v_1$ and $v_2$ can be expressed as linear combinations of the other two
 orthogonal vectors, $(\xi_2^*, \xi^*_1,0)^T$ and
 $(\xi^*_3 \xi_1, \xi^*_3 \xi_2, (|\xi_1|^2+|\xi_2|^2)\xi_3)^T$.

It is interesting to note that the combination: $N^{\prime 3}_R =
\xi_i N^i_R/\sqrt{\xi^*_j \xi_j}$ is simultaneously the eigenvector
of the zero eigenvalue of $\gamma$, $m$ and $Y'$. Choosing the other
two orthogonal combinations as:
\begin{eqnarray}
&&N^{\prime 1}_R = { \xi^*_2 N^1_R -\xi^*_1 N^2_R\over \sqrt{|\xi_1|^2+\xi_2|^2}}\;,\nonumber\\
&&N^{\prime 2}_R ={ \xi^*_3 \xi_1 N^1_R + \xi^*_3 \xi_2 N^2_R
-(|\xi_1|^2+|\xi_2|^2) N^3_R\over
\sqrt{(|\xi_1|^2+|\xi_2|^2)(|\xi_1|^2+|\xi_2|^2+|\xi_3|^2)}}\;,
\end{eqnarray}
and re-writing the Lagrangian ${\cal L}$ in terms of $N^{\prime
i}_R$, we find that $N^{\prime 3}_R$ decouples completely from the
theory. We have
\begin{eqnarray}
{\cal L} = \tilde \gamma_{ij}\bar N_R^{\prime i} \gamma_\mu (i
\partial^\mu N_R^{\prime j}) - {1\over 2} \left [\tilde m_{ij} \bar N_R^{\prime ic}
N^{\prime j}_R + 2 \tilde Y'_{ij} \bar L^i_L H N^{\prime j}_R + H.C.
\right ]\;,
\end{eqnarray}
where $\tilde \gamma$ and $\tilde m$ are now $2\times 2$ matrices,
and $\tilde Y'$ is a $3\times 2$ matrix.

One then further diagonalizes $\gamma = V^\dagger \hat \gamma V$ to
define new fields $\nu^{\prime}_R = V N^\prime_R$ and re-scale the
$\nu^{\prime i}_R$ fields by the square root values of the
eigenvalues of $\gamma$, $\gamma_i$, $\nu^i_R = \sqrt{
\gamma_i}\nu^{\prime}_R$. Finally one can rewrite the Lagrangian in
 the standard form
\begin{eqnarray}
{\cal L} = \bar \nu_R \gamma_\mu (i \partial^\mu \nu_R) - {1\over 2}
\left [ \bar \nu_R M \nu^c_R + 2 \bar L_L Y H \nu_R + H.C. \right
]\;,\label{final-form}
\end{eqnarray}
where $M$ is a $2\times 2$ matrix and $Y$  is a $3\times 2$ matrix.

Without the term proportional to $Y$, one can diagonalize $M$ to
reduce to Eq. (\ref{ff}). Actually even with non-zero $Y$, one can
still diagonalize $M = U^T \hat M U$ to have the first two terms in
the above equation look like Eq. (\ref{ff}), but the matrix $Y$
needs to be rotated with $\tilde Y = Y U^\dagger$.

The theory defined by the Lagrangian in Eq. (\ref{final-form}) is
identical to a theory of three left-handed and two right-handed
neutrinos, the minimal seesaw model~\cite{Frampton:2002qc}. The local FL
symmetry has reduced right-handed fields by one degree of freedom.

We comment that if the FL symmetry is a global one, there is no
constraint on the rank of the $\gamma_{ij}$ matrix. The linear
combination $N^{'3}_R$ does not disappear in the kinetic energy
terms. Only the mass matrix terms are affected. There is a massless
right-handed neutrino in the theory. This is the model considered in
Ref.~\cite{Jarlskog:2007qy,Jarlskog:2008zf}.

\vspace{0.3cm}
\noindent
{\bf Some implications}

We now discuss some implications of the model for right-handed
neutrinos to transform under a local FL transformation. After the
electro-weak symmetry breaking, that is the Higgs develops a
non-zero vacuum expectation value $<H> = v/\sqrt{2}$,
 the neutrino mass in the basis $(\nu_L, \nu_R^c)^T$
 is given by
 \begin{eqnarray}
 \left (\begin{array}{cc} 0&Y^*v/\sqrt{2}\\Y^\dagger v /\sqrt{2}&M \end{array} \right )\;.
 \end{eqnarray}
 This leads to the mass matrix $m_\nu$ for left-handed neutrinos to be
 \begin{eqnarray}
 m_\nu = - \frac{v^2}{2} Y^* M^{-1} Y^\dagger. \label{Numass}
 \end{eqnarray}
 One of the three light neutrinos has zero mass.

 It has been previously shown that the minimal seesaw model is consistent
 with experimental data~\cite{Frampton:2002qc,Barger:2003gt,xing10}, although
 the detailed numbers of data have changed~\cite{Schwetz:2008er}. We will not
 go into details about the phenomenology here, but would like to point out that
 the zero eigenvalue for the neutrino mass can be traced to the FL symmetry of the theory.

 Mathematically one understands why there is an zero eigenvalues
 by noting that $Y^\dagger$ is a $2\times 3$ matrix and is rank 2. It has an
 eigenvector with zero eigenvalue:
 \begin{eqnarray}
 Y^\dagger u_3 =0,~~m_\nu u_3 =0 \label{symm2}
 \end{eqnarray}
 Here $u_3$ is the vector introduced in Eq. (\ref{Yukaw1}).
 It is the eigenvector associated with the $v_3$ vector of the
 FL symmetry in the right-handed sector.

 We note that Eq. (\ref{symm2}) implies that after electro-weak
 symmetry breaking one gets a residual symmetry in the light neutrino
 mass term. The left-handed neutrinos in the mass term is invariant
 under the FL-like transformation,
 \begin{eqnarray}
 \nu \to \nu + u_3 z, \label{FLRes}
 \end{eqnarray}
 where $\nu = (\nu_L^1, \nu^2_L, \nu_L^3)^T$.
 We start with a FL symmetry, Eq. (\ref{FL2}), of the full Lagrangian
 and end up with a residual FL symmetry for the seesaw masses of
 neutrinos. Note that the original FL symmetry applies to the
 right-handed neutrinos and the residual FL symmetry applies to
 left-handed neutrinos which can be traced back to the requirement that $y'_3 =0$
 in Eq. (\ref{Yukaw1}) dictated by the FL symmetry. The zero mass of a light
 neutrino is therefore a consequence of the FL symmetry. If the
 transformation is global, then this residual symmetry also applies
 to the kinetic energy terms.

It is interesting to note that any mass matrix for fermionic field
$\nu$ with a zero eigenvalue, one can define a FL-like
transformation related to the associated eigenvector $u$: $\nu \to
\nu + u z$. Under this transformation, the mass term is invariant.


If future experimental data will determine that all three light
neutrinos to have non-zero masses, the minimal seesaw needs to be
extended. One might wonder if higher order loop corrections can make
all three light neutrino masses non-zero. We find that this is not
true because in the theory the FL is not broken, the masslessness of
one of the neutrinos is true to all orders. To obtain a theory with
at least three non-zero mass light neutrinos with FL symmetry
imposed on a particular direction in $\xi_i$ parameter space, more
fields need to be introduced. In our case since the local FL
symmetry always reduce the number of fields by one, we need to start
with more than three right-handed neutrinos. For example, starting
with 4 right-handed neutrinos, after the reduction discussed before,
the $M$ and $Y$ matrices in Eq. (\ref{ff}) become $3\times 3$
matrices. The resulting theory is a (3+3) seesaw model.

\vspace{0.3 cm}
\noindent
{\bf Conclusions}

 In summary we have studied consequences of the Friedberg-Lee symmetry
 for seesaw models. We find that if a local FL symmetry
 is imposed to the full Lagrangian of right-handed neutrinos,
 one of the right-handed neutrinos completely decouples from the theory.
 For specific model studies, we begin with a $3+3$ seesaw model, which is a model with
 three generations of left-handed and right-handed neutrinos. After applying a local FL symmetry
 to the right-handed sector, we arrive at
 a $3+2$ seesaw model, the minimal seesaw model, which is a model with three generations of
 left-handed neutrinos and two generations of right-handed neutrinos.
 In this model one of the light neutrinos has
 zero mass as a consequence of the FL symmetry. The masslessness of one
 light neutrino means that there is a FL symmetry in the seesaw
 mass matrix of the light left-handed neutrinos. This FL symmetry
 in the seesaw mass matrix of the light left-handed neutrinos is a consequence of the
 FL symmetry imposed on right-handed sector of neutrinos in the original
 seesaw model. The FL symmetry can  enhance the predictive power of a theory.

\acknowledgements
 We thank Professor A. Zee for many useful discussions.
 This work is supported in part by NSC and NCTS, and the Science and Technology Commission of
 Shanghai Municipality under contract number 09PJ403800.


\end{document}